\documentclass[aps,pra,reprint,showpacs,superscriptaddress]{revtex4-1}

\usepackage{graphicx}
\usepackage{hyperref}
\usepackage{amssymb,amsmath}
\usepackage{color}
\usepackage{mathrsfs}

\newcommand{\eqr}[1]{Eq.~(\ref{#1})}

\newcommand{\figr}[1]{Fig.~\ref{#1}}

\newcommand{\refr}[1]{Ref.~\cite{#1}}

\newcommand{\braca}[1]{\left(#1\right)}
\newcommand{\bracb}[1]{\left[#1\right]}
\newcommand{\bracc}[1]{\left|#1\right|}

\newcommand{\ket}[1]{|#1\rangle}

\usepackage{calligra} \DeclareMathAlphabet{\mathcalligra}{T1}{calligra}{m}{n} \DeclareFontShape{T1}{calligra}{m}{n}{<->s*[2.2]callig15}{}   

\begin{document}

\title{Single-atom single-photon coupling facilitated by atomic-ensemble dark-state mechanisms}
\author{Andrew~C.~J.~Wade}
	\affiliation{Department of Physics and Astronomy, Aarhus University, Ny
  Munkegade 120, DK-8000 Aarhus C, Denmark.}
\author{Marco Mattioli}
	\affiliation{Institute for Theoretical Physics, University of Innsbruck, A-6020 Innsbruck, Austria}
	\affiliation{Institute for Quantum Optics and Quantum Information of the Austrian Academy of Sciences, A-6020 Innsbruck, Austria}
\author{Klaus M\o{}lmer}
	\affiliation{Department of Physics and Astronomy, Aarhus University, Ny
  Munkegade 120, DK-8000 Aarhus C, Denmark.}

\date{\today}

\begin{abstract}
We propose to couple single atomic qubits to photons incident on a cavity containing an atomic ensemble of a different species that mediates the coupling via Rydberg interactions.
Subject to a classical field and the cavity field, the ensemble forms a collective dark state which is resonant with the input photon, while excitation of a qubit atom leads to a secondary "dark" state that splits the cavity resonance. The two different dark state mechanisms yield zero and $\pi$ reflection phase shifts and can be used to implement quantum gates between atomic and optical qubits.
\end{abstract}

\maketitle

\section{Introduction}

Numerous proposals for large scale quantum computing and communication employ coupling of flying and stationary qubits in the form of photons and atoms \cite{kimble2008a}. Optically dense media allow efficient protocols for coupling light to collective qubit degrees of freedom in atomic samples \cite{duan2001a}, while a high Q cavity may provide strong coupling between a single photon and a single atom \cite{duan2004}, allowing non-destructive detection of the reflection of a single photon by its modification of an atomic superposition state \cite{reiserer2013}. Recently, these efforts culminated with the demonstration of deterministic entanglement between a single atom and photons reflected from a high Q cavity, \cite{Reiserer2014a}.

Recent experiments have demonstrated single-photon phase modulation due to traversal of an atomic medium with a stored Rydberg polariton \cite{Beck2015a}, and in \refr{Hao2015a}, a phase gate was proposed where a light pulse scattering on a cavity that contains a qubit stored in an atomic ensemble acquires a conditional reflection phase shift. Similarly, \refr{Das2016a} proposed a photonic phase gate between photons sequentially scattered under electromagnetically induced transparency (EIT) conditions, where the first photon is stored in a collective Rydberg state, taking the place of the atomic qubit in \refr{Hao2015a}. In this article, we propose to mediate an effective single-atom single-photon interaction by employing individually trapped qubit atoms and an atomic ensemble of a different species. By the inter-species dipolar interaction a single Rydberg excited qubit atom can couple strongly to the Rydberg states of the surrounding ensemble atoms \cite{Mueller2009,Saffman2009a}. This, in turn, affects their collectively amplified coupling to light fields \cite{Lukin2001a,Saffman2002a,Pedersen2009a,ningyuan2016a,Ding2015a} and thus the optical properties of the system at the single photon level.

In our proposal, the qubit information is retained in single register atoms and photons while the ensemble only serves to mediate an effective, collectively enhanced and broad bandwidth interaction between them. In a quantum repeater architecture, the effective coupling to light mediated by the ensembles applies also for repeater nodes containing several individual qubit atoms and paves the way for deterministic quantum gates and distillation protocols. Elements of our proposal may be demonstrated in single species experiments by exciting the qubit and the ensemble atoms to different Rydberg states as in \refr{Das2016a}. Compared to the dual species implementation, however, maintaining the separate roles as qubit and ensemble atoms puts stronger requirements on the spatial separation and addressing of the otherwise indistinguishable ground state atoms. Our use of different atomic species makes the addressing much easier and, importantly, it implies that the light interacting with the ensemble is not resonant with any transition in the qubit atoms. The qubit information may thus be protected from ensemble decoherence mechanisms, \cite{Derevianko2015a,Goldschmidt2016a}, and does not degrade due to re-absorption of fluorescence emitted from the ensemble \cite{Beterov2015a}.

The article is organized as follows: In Sec. II, we present the atomic level scheme and our use of single and two-atom adiabatic dark states in our gate protocols. In Sec. III, we provide a quantitative input-output analysis of the single photons scattering on the cavity holding the atomic system. In Sec. IV, we describe applications for atom-photon and atom-atom phase gates. In Sec. V, we present numerical analyses of the gate fidelities using real atomic parameters and realistic assumptions for the optical cavity and the trapped atomic ensembles. Sec. VI concludes the article.

\section{Photon phase shifts due to atomic dark state dynamics}

\begin{figure}[ht]
  \centering
 \includegraphics[width=.48\textwidth]{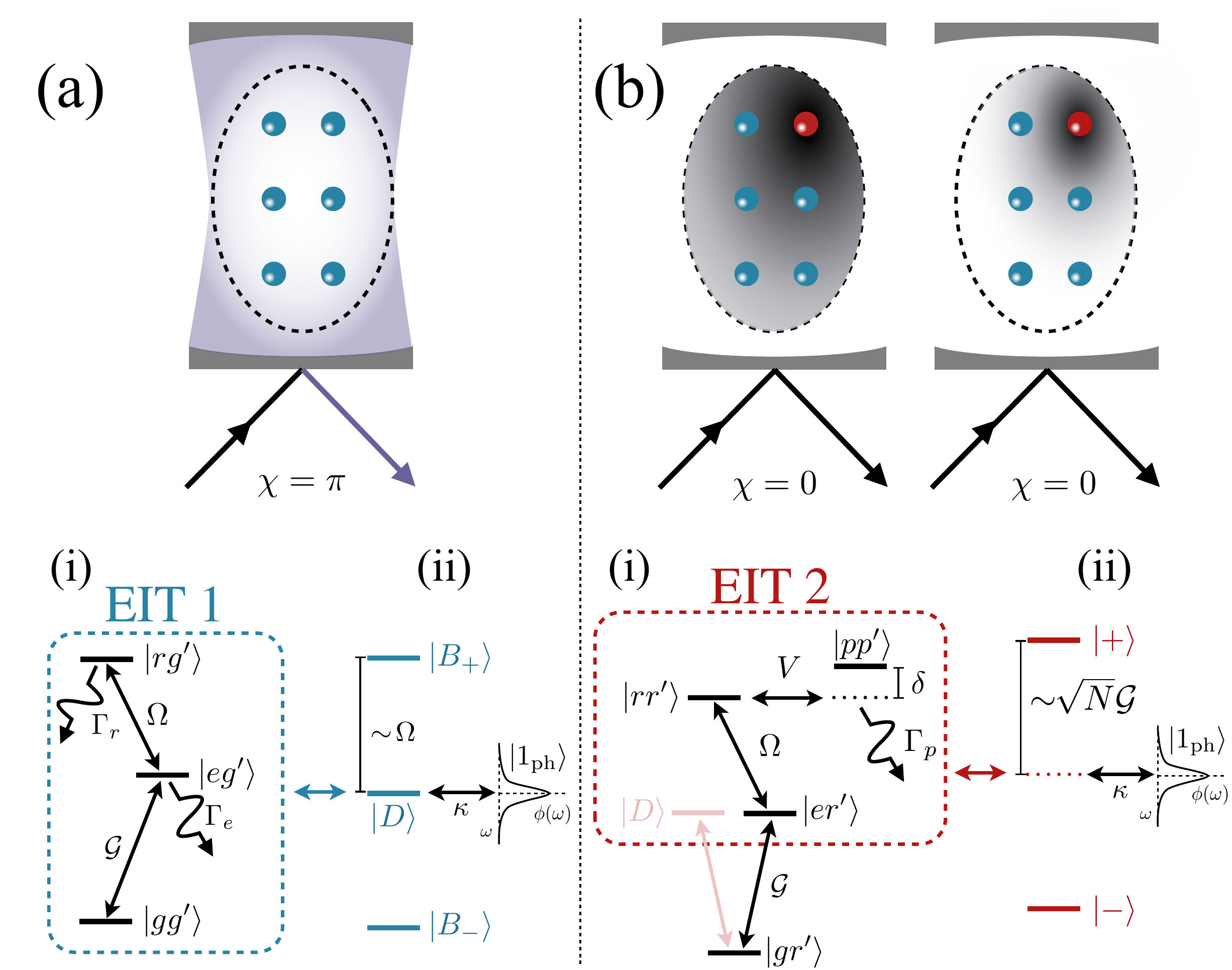}
  \caption{(Color online) Panels (a,b) depict Cs atoms (blue and red circles) within in an ensemble of Rb atoms (dashed oval). Panel (a) shows how an incident photon couples into an EIT dark state $\ket{D}$ and exits with the phaseshift $\chi\sim\pi$. (a,i) the Rb ground state $\ket{g}$, excited state $\ket{e}$ (decay $\Gamma_e$) and Rydberg state $\ket{r}$ (decay $\Gamma_r$) energy levels near a ground state Cs atom $\ket{g'}$ are coupled by the (quantum) cavity field and a laser control field. Panel (b) shows the effect of a Cs atom excited to state $\ket{r'}$ where the state $\ket{r,r'}$ is nearly resonant with a  F\"orster coupled state $\ket{p,p'}$ (b,i). The dipolar coupling $V$ and the laser field forms the "EIT 2" configuration with two "bright" states (not shown) and a "dark" state ($\ket{D}$ faded red), resonant with the cavity field. The collective coupling to the Rb ensemble splits the cavity resonance and causes the reflection of the incident field with the phase shift $\chi\sim 0$. All, or only nearby, Rb atoms may form the EIT 2 dark state, as depicted by the grey shading in (b).\label{fig:system_cartoon}}
\end{figure}

We shall analyze our proposal for the special case of cesium (Cs) register and rubidium (Rb) ensemble atoms, which have favourable interactions among specific Rydberg states \cite{Beterov2015a}. In \figr{fig:system_cartoon}, panels (a,b) show a schematic set-up with blue (red) balls representing Cs ground state (Rydberg excited) atoms and the dashed oval representing the circumference of the ensemble of $N$ Rb atoms.

Panel (a) assumes the Cs ground state $\ket{g'}$, where (a,i) shows the Rb level scheme with the ground state $\ket{g,g'}$ coupled by the quantized cavity field to the excited state $\ket{e,g'}$ with (single-atom) coupling strength $\mathcal{G}$. A resonant laser, coupling the Rydberg level $\ket{r,g'}$ with strength $\Omega$ to $\ket{e,g'}$, results in the, so-called, EIT dark state $\ket{D}\sim \Omega \ket{g,g'}-\mathcal{G}\ket{r,g'}$ \cite{Fleischhauer2005a,Pritchard2010a} with a vanishing state $\ket{e,g'}$ amplitude due to destructive interference, and the two bright states shown in (a,ii).

Panel (b) assumes a Cs atom is excited to a Rydberg state $\ket{r'}$, where (b,i) shows the Cs-Rb pair of states $\ket{r,r'}$ is coupled by a dipole-dipole interaction with the strength $\hbar V$ to a nearly degenerate pair of Rydberg states $\ket{p,p'}$ (energy defect $\hbar\delta$). If $\delta\simeq 0$, the states $\ket{e,r'}$, $\ket{r,r'}$ and $\ket{p,p'}$ form a correlated two-atom EIT configuration (EIT 2), leading to a "dark" state $\sim V\ket{e,r'}-\Omega \ket{p,p'}$ with no $\ket{r,r'}$ amplitude (b,i, faded red) and two "bright" eigenstates (not shown). The bright states are shifted in energy, while the dark state can be excited resonantly from $\ket{g,r'}$ by absorption of a cavity photon. This coupling is collectively enhanced by the presence of $N$ Rb atoms and results in normal-mode splitting, shifting the incident photon out of resonance by $\pm\sqrt{N}\mathcal{G}$ (b,ii).

The photon is ultimately reflected from the cavity, but the intermediate excitation of the dark state in panel (a) causes a $\pi$-phase shift compared to the reflection from the off-resonant cavity in panel (b).
The Cs atom hyperfine ground state provides qubit states $\ket{g'_0}\equiv \ket{0}$, $\ket{g'_1}\equiv\ket{1}$, and by resonant excitation of Cs from $\ket{0}$ to $\ket{r'}$, the phase of the photon is controlled by the qubit state of a single Cs register atom, but benefiting from the collectively enhanced coupling strength with the $N$-atom Rb ensemble.

\section{Phase shifts from input-output theory}

In our quantitative analysis, detailed in the Appendix, we apply the formalism of \cite{Das2016a}, and we obtain the complex reflection coefficient for a photon incident on the cavity\begin{align}
R_{\{j\}}(\omega)= 1{}&{} - \kappa \left\{\frac{\kappa}{2} - i \omega + \sum_{n=1}^N  |\mathcal{G}_n|^2 \left[ \frac{\Gamma_{e}}{2}-i\omega \phantom{+\frac{|\Omega|^2}{\frac{\Gamma_{r}}{2} - i \omega + \frac{|V_{jn}|^2}{\frac{\Gamma_p}{2}+i(\delta-\omega)}}}\right .\right . \nonumber \\
&\left.\left.\qquad\qquad{}+\frac{|\Omega|^2}{\frac{\Gamma_{r}}{2} - i \omega + \frac{\sum_{\{j\}} |V_{jn}|^2}{\frac{\Gamma_p}{2}+i(\delta-\omega)}}\right]^{-1}\right\}^{-1},\label{eq:full_R}
\end{align}
(where $\{j\}$ denotes the set of excited Cs atoms). In our calculation, we assume that the cavity decoherence is dominated by the mirror transmission loss
and we have, for simplicity, omitted the decay of the Cs Rydberg excited states at this stage. In the absence of Rb-Rb Rydberg interactions, the optical reflection process is linear, and the coefficient (\ref{eq:full_R}) relates the quantized field operators in the input-output formalism, assuming an additional Langevin noise term to ensure unitarity \cite{gardiner2004a}. We have not, however, made any assumptions about the interactions among Rydberg excited Rb atoms, as we shall apply Eq.(\ref{eq:full_R}) only for the case where the cavity is illuminated by a single photon and, hence, at most a single Rb atom is excited (see Appendix).

We recall, that when no Cs atoms is excited ($\{j\}=\varnothing$), we want $R_{\varnothing }(\omega)=-1$, and if one or several Cs atoms are excited, we want $R_{\{j\} \neq \varnothing }(\omega)=1$, over the frequency range $\Delta_\omega$ spanning the incident photon spectrum. This is accomplished due to the interaction term in the second line of \eqref{eq:full_R} which has a significantly stronger effect for our F\"orster coupled Rydberg states than in the analyses \cite{Hao2015a,Das2016a} applying Rydberg (blockade) energy shifts.
\subsection{Performance estimates}

We shall apply a straightforward numerical evaluation of \eqr{eq:full_R} for realistic spatial configurations of the atoms, but to gain semi-quantitative insight in the performance of our coupling scheme as function of different parameters, we shall here approximate the system by separating the contributions to \eqr{eq:full_R} from ensemble atoms that interact and do not interact with an excited qubit atom. Taking a homogenous cavity coupling $\mathcal{G}_n\rightarrow \mathcal{G}$, and applying $V_{jn} = C_3/r_{jn}^3$ where $r_{jn}$ is the distance between the Cs and Rb atoms, \eqr{eq:full_R} suggests that Rb atoms are unperturbed by Cs atoms beyond the distance $R_B=(\Gamma_e \bracc{C_3}^2/\gamma |\Omega|^2)^{\frac{1}{6}}$ where $\gamma = \mathrm{max}(|\delta|,\Gamma_p)$.
Assuming that the interaction term dominates the denominator of the second line of \eqr{eq:full_R} for a homogeneous density of Rb atoms within $R_B$ of the Cs atom, \eqr{eq:full_R} simplifies to
\begin{equation}
R(\omega)= 1 - \braca{\frac{1}{2}-i\frac{\omega}{\kappa} + f_B(\omega)  N_B C  + f_E(\omega) N_{E}C }^{-1},\label{eq:R_simple}
\end{equation}
where $N_B$  ($N_E = N-N_B$) are the number of atoms perturbed (unpertubed) by the Cs atoms,
$C = |\mathcal{G}|^2/\kappa\Gamma_e$ is the single atom cooperativity, and integrals within and beyond $R_B$ yield

\begin{eqnarray} \label{eq:fB}
f_B(\omega) =-(1+i)\tan^{-1}\bracb{\frac{(1+i)\alpha}{\sqrt{i+2\omega/\Gamma_e}}}/\alpha\sqrt{i+2\omega/\Gamma_e}
 \end{eqnarray}
 with $\alpha=\left[\frac{\Gamma_p}{2\delta}+i(1-\frac{\omega}{\delta})\right]^\frac{1}{2}$, and,
\begin{eqnarray} \label{eq:fE}
f_E(\omega)=i\Gamma_e\left(\frac{i\Gamma_e}{2}+\omega-\frac{2|\Omega|^2}{2 \omega + i \Gamma_r}\right)^{-1},
\end{eqnarray}
respectively.

When one or several Cs atoms are excited to the Rydberg state, we want the term $ f_B(\omega) N_B C$ to be larger than all other terms in the parenthesis in \eqr{eq:R_simple}, such that we obtain $R(\omega)\sim 1$ for the bandwidth $\Delta_\omega$. For $\omega=0$ and $\Gamma_p \ll \delta$, $f_B\sim i-1$, and we request that the frequency splitting is larger than the cavity linewidth $\sqrt{N_B}|\mathcal{G}| > \kappa$, equivalent to the request that the cooperativity of the perturbed part of the ensemble $N_B C \gg 1$. For the case of no Cs Rydberg excitation, the EIT mechansim should yield $R(\omega)\sim -1$ around resonance. The last term $f_E(\omega) N_E C$, however, represents a phase shift and decoherence associated with the atomic excited state component of the EIT dark state. For $\Gamma_r=0$, the resulting $|R(\omega)|^2$ becomes Lorentzian with a width $\frac{|\Omega|^2}{\sqrt{N}|\mathcal{G}|}\sqrt{\frac{\kappa}{\Gamma_e}}$.

Our simple estimates are confirmed by numerical calculations and they show that the reflection phase shift can be accurately controlled by the Cs qubit atoms. We shall now present a few applications for quantum information processing tasks, obtaining expressions for their fidelities that we then compute for realistic experimental situations by a numerical treatment of \eqr{eq:full_R}.

\section{Application for atom-photon and atom-atom phase gate}

Rather than merely scattering a single photon on the cavity, we may employ a photonic qubit, with a two-dimensional Hilbert space associated with either a zero and a one photon state component, or with (dual rail) spatial, time-bin, or frequency components. If only one of the photonic qubit components experiences the reflection phase shift controlled by the Cs qubit levels, we obtain a single-atom single-photon phase gate. The fidelity of the gate is reduced if the reflection is imperfect and if the photon occupies different mode functions before and after the reflection. The distortion of a normalized temporal mode function $\phi(t)$ ($\int |\phi(t)|^2 dt = 1$) is readily obtained in frequency space, where the mode amplitude $\phi(\omega)\rightarrow R_{\{j\}}(\omega)\phi(\omega)$, with
$R_{\{ j\}}(\omega)$ given in \eqr{eq:full_R}. The resulting loss of fidelity of the phase gate depends on the input qubit states and on the overlap functions,
\begin{eqnarray} \label{eq:Tj}
T_{\{ j\}} = \int_{-\infty}^{\infty} d\omega \bracc{\phi(\omega)}^2 R^*_{\{ j\}}(\omega).
 \end{eqnarray}
 Following arguments given in \cite{Das2016a} we obtain the average fidelity over all possible input atom and photon qubit states,
\begin{equation}
F_{\mathrm{at \cdot ph}} = \frac{1}{16}\bracc{2+T_{\{ j\}} - T_{\{ \varnothing \}}}^2. \label{eq:Fatph}
\end{equation}
with numerical values depicted in \figr{fig:fidelity}(a,d,g).

\begin{figure}[h]
  \centering
 \includegraphics[width=.48\textwidth]{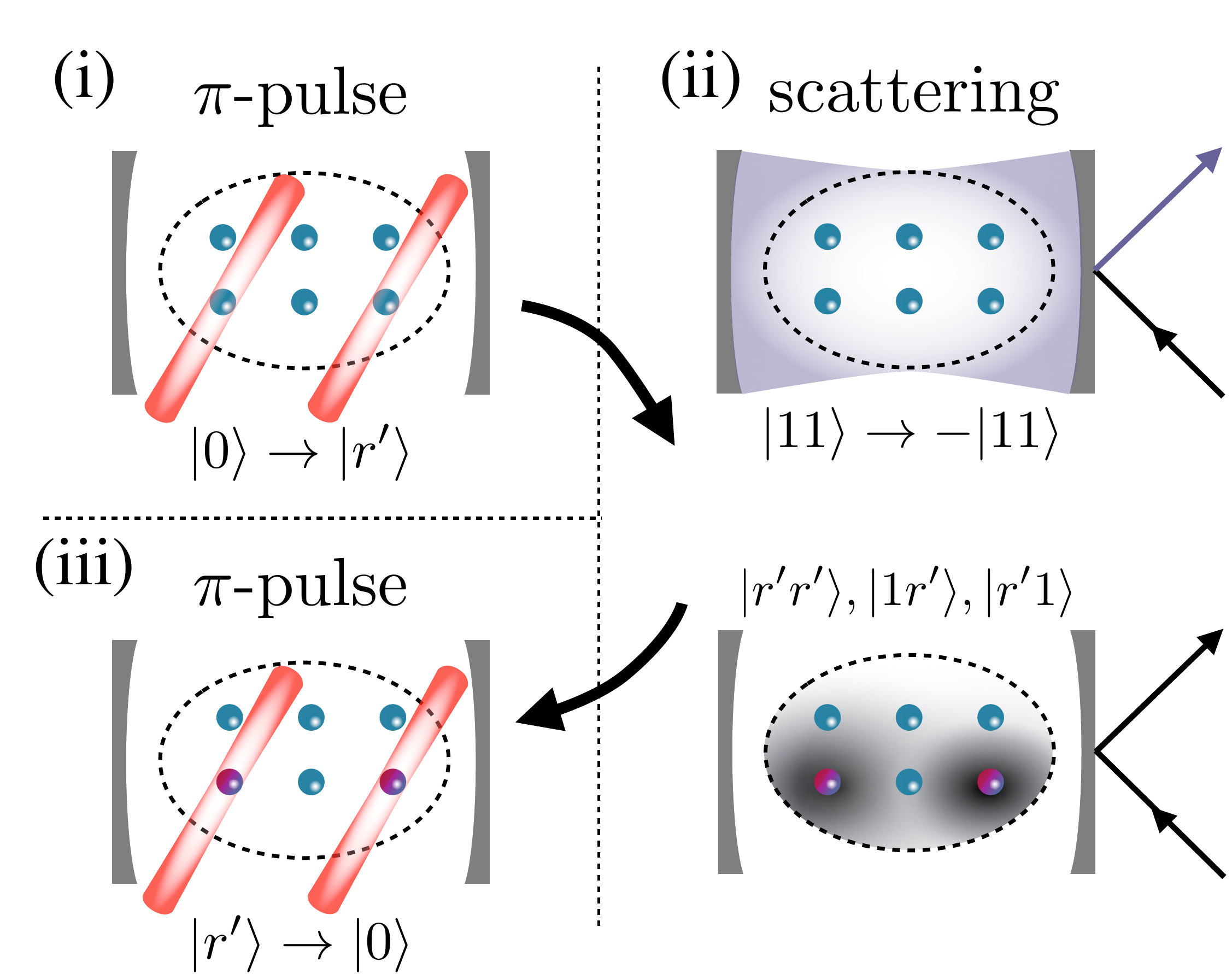}
  \caption{(Color online) A controlled-phase gate between two non-interacting Cs atoms: (i) the atoms are excited from $\ket{0}$ to $\ket{r'}$. (ii) a single photon is scattered off the cavity and $\ket{11}$ undergoes a sign change with respect to the other states. (iii) a $\pi$-pulse returns the excited atoms from $\ket{r'}$ to  $\ket{0}$.\label{fig:pi_phase}}
\end{figure}

The phase shift accompanying reflection of a photon with no qubit degrees of freedom has interesting applications on systems with several qubit atoms. Consider the situation sketched in \figr{fig:pi_phase}, where we assume that the Cs qubit atoms are far apart and do not interact with each other. If none of the cesium atoms are in the Rydberg state, the photon is scattered via the Rb dark state and undergoes a $\pi$-phase shift, while if one, or several cesium atoms occupy their Rydberg states, the surrounding rubidium ensemble causes the mode splitting and reflection of the incident photon with a vanishing phase shift. The reflection phase factor on the photon thus acts to produce different phases on different states of the cesium register, as illustrated with the implementation of an atom-atom phase-gate in \figr{fig:pi_phase}. The fidelity of this gate depends on the distinguishability of the reflected photon wave packets for the different atomic qubit states. A simple calculation shows that the average fidelity over all input atomic qubit states becomes
\begin{equation}
F_{\mathrm{at \cdot at}} = \frac{1}{16}\sum_{\{j\},\{k\}}\Theta(\{j\}) \Theta(\{k\}) T^{\{j\}}_{\{k\}}. \label{eq:Fatat}
\end{equation}
where $\{j\}$ and $\{ k\}$ explore the four combinations of atoms that are either excited from $\ket{0}$ to $\ket{r'}$ or remain in the qubit state $\ket{1}$, $\Theta(\{j\}=\varnothing) = -1$, $\Theta(\{j\}\neq \varnothing) = 1$, and
\begin{eqnarray} \label{eq:Tjk}
T^{\{ j\}}_{\{ k\}} = \int_{-\infty}^{\infty} d\omega \bracc{\phi(\omega)}^2 R_{\{ j\}}(\omega) R^*_{\{ k\}}(\omega).
\end{eqnarray}
Numerical values for $F_{\mathrm{at \cdot at}}$ are depicted in \figr{fig:fidelity}(b,e,h).

\section{Numerical analysis}

For a quantitative assessment of the atom-photon and atom-atom phase gates, we consider the recent experimental set-ups with densities $10^{11}-10^{12}\,\mathrm{cm^{-3}}$ \cite{Parigi2012a} to $10^{14}-10^{15}\,\mathrm{cm^{-3}}$ \cite{Brennecke2007a,Murch2008a}. A Gaussian Rb density distribution $n=N\exp[-(x^2+z^2)/R_c^2 - y^2/R_y^2]/(R_c^2 R_y\pi^{3/2})$ with $(R_c,R_y,R_\mathcal{G})=(5,20,15)\mathrm{\mu m}$ overlaps a standing wave cavity field $\mathcal{G}_n=\mathcal{G}_0 \sin (\frac{2\pi}{\lambda} y_n) \exp[-(x_n^2+z_n^2)/R_\mathcal{G}^2]$ addressing the $\lambda=788$ nm  D$2$ transition ($\Gamma_e=3 $MHz) with $\mathcal{G}_0 \simeq 1$ MHz.

We study a photon-atom phase gate with the Cs qubit atom located in the middle of the Rb cloud $y=0\mathrm{\mu m}$, and an atom-atom phase gate between Cs atoms positioned along the principal axis of the Rb distribution at  $y=\pm15\mathrm{\mu m}$. The Cs atoms may be held in position by optical potentials which may be designed to repel Rb atoms and thus suppress collisions that could decohere the qubit information.
The Rydberg Cs-Rb interaction is
\begin{equation}
V_{jn}(\theta_{jn},r_{jn})=\sqrt{f(\theta_{jn})}  \frac{C_{3}}{r_{jn}^3}, \nonumber
\end{equation}
where two prominent F\"{o}rster resonances are ($F1$) Rb48s$_{1/2}$Cs51s$_{1/2}$ $\leftrightarrow$ Rb48p$_{3/2}$Cs50p$_{1/2}$, $C^{F1}_{3}=1.69$GHz $\mu m^3$, and $\delta_{F1}=-5.71$MHz, and, ($F2$) Rb84s$_{1/2}$Cs89s$_{1/2}$ $\leftrightarrow$ Rb84p$_{1/2}$Cs88p$_{1/2}$, $C^{F2}_{3}=-18.2$GHz $\mu m^3$, and $\delta_{F2}=-2.43$MHz \cite{Beterov2015a,SaffmanPa}.
For $F1$ with Cs aligned and Rb anti-aligned along the $z$-axis $f(\theta_{jn}) = [10+6\sin^2(\theta_{jn})]/9$, while for $F2$ with aligned Cs and Rb, $f(\theta_{jn}) = [4+6\sin^2(\theta_{jn})]/9$, where $\theta_{jn}$ is the polar angle of the relative vector.

\begin{figure}[th]
  \centering
  \includegraphics[width=.48\textwidth]{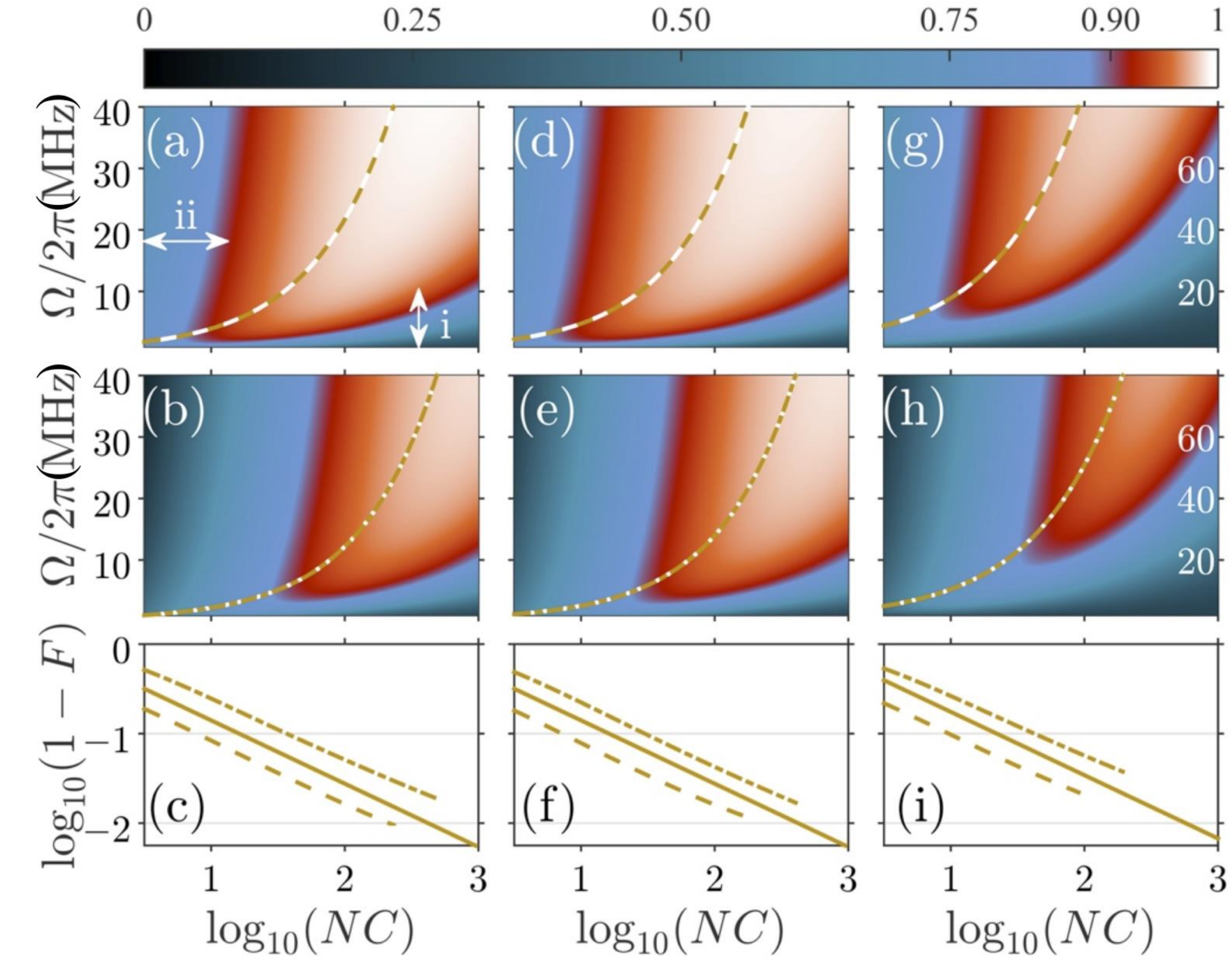}
  \caption{(Color online) The dependence of $F_{\mathrm{at \cdot ph}}$ (top row) and $F_{\mathrm{at \cdot at}}$ (middle row) on the laser field coupling strength $\Omega$ and on the total number of atoms multiplied with the single atom cooperativity, $NC$, with photon bandwidth $\Delta_\omega= 10$ kHz and cavity linewidth  $\kappa= 10$ MHz for the F\"{o}rster resonances $F2$ (left column) and $F1$ with $\delta_{F1}=0$ (middle column). The same structure is observed with higher values of $\Omega$ ($y$-axis is multiplied by $2$) in the right column for $F2$ with $\delta_{F2}=0$ and $(\Delta_\omega,\kappa)= (1,30)$ MHz. The bottom row shows the infidelity minimized over $\Omega$ for $F_{\mathrm{at \cdot ph}}$ (dashed) and $F_{\mathrm{at \cdot at}}$ (dotted) [scales as $\sim 1/\sqrt{NC}$ (solid)], cf.,  the gold lines in the corresponding $2$D plots. \label{fig:fidelity}}
\end{figure}

For $F2$, the dependence of $F_{\mathrm{at \cdot ph}}$ [\figr{fig:fidelity}(a)] and $F_{\mathrm{at \cdot at}}$ [\figr{fig:fidelity}(b)] on the applied laser field coupling strength $\Omega$ on the Rb $\ket{e}\rightarrow\ket{r}$ transition with $\Gamma_r=\Gamma_p= 10 $kHz, and on the total number of atoms through the collective  cooperativity, $NC$, is shown.
We assume Gaussian photon pulses with a frequency profile $\phi(\omega)=\exp[-\omega^2/(2 \Delta_\omega^2)]/\sqrt{\sqrt{\pi} \Delta_\omega}$ with $ \Delta_\omega= 10$ kHz.
Moving along arrow $i$ in \figr{fig:fidelity}(a), the dominant source of infidelity is the phase shift and decoherence associated with the unperturbed Rb atoms, $f_E(\omega) N_E C$ [c.f., \eqr{eq:R_simple}], being small for $N C  \ll |\Omega|^2 /\Gamma_e \Gamma_r$. Conversely, the dominant source of infidelity along arrow $ii$ is the unwanted excitation of the cavity mode due to the collective shift $\sqrt{N_B} |\mathcal{G}_0|$ not exceeding linewidths, i.e., $N_B C\sim1$. As the distance $R_B$, within which the Rb atoms are perturbed by the presence of a Rydberg Cs atom, depends on $\Omega$, and $N_B\sim N R_B/R_y$ in the elongated geometry, we require $N C\gg R_y/R_B \propto |\Omega|^{1/3}$ for small infidelity.

Rydberg Rb-Rb interactions are not present as there is only a single Rydberg Rb excitation via the EIT dark state when no Cs atoms are excited. However, in the case of the atom-atom gate, the two Rydberg Cs atoms have an interaction energy of $U/\hbar=(0.02,20)$kHz for (Cs51s, Cs89s), which will reduce the fidelity if we use the highly excited Cs89s.

This interaction can be reduced via other geometries, e.g., larger separations where each Cs atom is surrounded by a small cloud of Rb atoms \cite{Komar2016a}, or by using the lower Rydberg Cs state Cs51s. The latter state has a much weaker interaction with the Rb ensemble, but by tuning the Rydberg levels, e.g., with ac \cite{Bohlouli2007a} or dc \cite{Vogt2006a,Urban2009a} electromagnetic fields, the energy defect $\delta_{F1}$ of the F\"{o}rster resonance can be reduced to zero, and thus, we maximize the effect of the interaction term $V_{jn}$ in \eqr{eq:full_R}.

The fidelity $F_{\mathrm{at \cdot ph}}$ [$F_{\mathrm{at \cdot at}}$] for $F1$ with $\delta_{F1}=0$ is shown in \figr{fig:fidelity}(d) [\figr{fig:fidelity}(e)] and is found to be quantitatively similar to $F2$ without tuning, \figr{fig:fidelity}(a) [\figr{fig:fidelity}(b)], owing to $[C^{F2}_{3,1}]^2/\delta_{F2} \sim [C^{F1}_{3,3}]^2/\Gamma_p$ [c.f., \eqr{eq:full_R}].
Additionally, the communication bandwidth can be increased by reducing the temporal extent of the photon, validating the neglect of the Cs-Cs interaction and Cs Rydberg state decay. To maintain high fidelities, we then require an increased $\Omega$ (to increase the EIT linewidth and reduce the phase shift and decoherence) and an increased cavity linewidth $\kappa$, and, consequently, increased $N$.
For $\sim 1\mathrm{\mu s}$ photons ($\Delta_\omega=$ MHz), $F_{\mathrm{at \cdot ph}}$ [$F_{\mathrm{at \cdot at}}$] is shown in \figr{fig:fidelity}(g) [\figr{fig:fidelity}(h)], with the high fidelity regimes shifted to higher values of $\Omega$ and $NC$.
In panels (c,f,i) of \figr{fig:fidelity}, we observe a $\sim 1/\sqrt{NC}$ scaling of the infidelity with the collective cooperativity, minimized over $\Omega$.

\section{Conclusion}

To summarize, we have shown that the collective coupling of an ensemble of atoms to a cavity mode together with a novel two-atom EIT coupling mechanism among Rydberg states can be employed for phase gates between photons and single atoms. Using parameters corresponding to current cavity QED experiments with atomic ensembles we find atom-photon gate fidelities which are adequate for quantum communication purposes. The atom-photon gate also allows dispersive readout of individual atomic qubits with similar suppression of cross-talk as a recent proposal \cite{Beterov2015a} using Rydberg interactions to obtain a fluorescence signal from an auxiliary atom.

Our ability to use photonic reflection phase shifts to perform atom-atom (and multi-atom \cite{Molmer2011a}) gates brings promise for applications such as  entanglement distillation, decoherence-free subspace encoding and error correction algorithms in few-qubit registers. As the controlled atom-photon phase gates can be applied successively on several cavities, our scheme may find applications in strategies for distributed quantum computing as well as for the establishment of networks of entangled states with applications to distributed clocks and high precision measurements.
\newline
\begin{acknowledgments}
This work was supported by the ARL-CDQI program through cooperative agreement W911NF-15-2- 0061, the EU H2020 FET-Proactive project RySQ, the SFB FoQuS (FWF Project No. F4016-N23) of the Austrian Science Fund, the Marie Curie Initial Training Network COHERENCE, and the Villum Foundation. The authors acknowledge D. Petrosyan, M. Saffman, P. Zoller, B. Vermersch and A. Glaetzle for fruitful discussions and comments on the manuscript.
\end{acknowledgments}

\appendix*

\section{} \label{app_refl}

Here, we shall calculate explicitly the reflection coefficient $R_{\{j\}}(\omega)$ in Eq.~\eqref{eq:full_R} of the main text.

For notational simplicity we assume that there is only one Cs atom and omit the label $\{j\}=1$ of the main text; the final result is readily extended to the case of many Cs atoms.
We first define the total Hamiltonian

\begin{align} \label{HAM}
H = &\,H_{\mathrm{sys}} + H_{B_{c}}+ H_{\mathrm{int}, B_c} + H_{B_{a}}+ H_{\mathrm{int}, B_a},
\end{align}
where the system Hamiltonian models a cavity that contains a Cs atom and an ensemble of $N$ Rb atoms,
\begin{align} \label{sysham}
H_{\mathrm{sys}} = &\sum_{n=1}^N  \left( -\hbar \Omega |r_{n}\rangle\langle e_{n}|-i\hbar \, \mathcal{G}_{n}|e_{n}\rangle\langle g_{n}|\hat{b}\right) +\text{H.c.} + \nonumber \\ + &\hbar \sum_{n=1}^N V_{n}|r_{n}\rangle\langle p_{n}|\otimes|r^{\prime}\rangle\langle p^{\prime}| + \text{H.c.} +\nonumber \\
+ &\hbar \, \delta \sum_{n=1}^N |p_{n}\rangle\langle p_{n}|\otimes|p^{\prime}\rangle\langle p^{\prime}|
\end{align}
(in the interaction picture with respect to the cavity frequency $\omega_c$). $\mathcal{G}_n$ is the coupling of the single-mode cavity field, $\hat{b}$ ($[\,\hat{b},\hat{b}^{\dagger}] = 1$), with the transition from the ground-state $|g_n\rangle$ to the excited state $|e_n \rangle$ of Rb atom $n$ of the ensemble. The latter is at the same time driven by a strong and homogeneous classical field with Rabi frequency $\Omega$ to the Rydberg state $|r_n\rangle$. We assume vanishing single- and two-photon detunings $\Delta = \omega_c -\omega_{eg}$ and $\delta^{\prime} = (\omega_c - \omega_{eg}) +  (\omega_{Lre} - \omega_{re})$, where, $\hbar \omega_{eg} (\hbar \omega_{re})$ is the energy of the excited (Rydberg) state with respect to the energy of the ground (excited) state, while  $\omega_{Lre}$ is the frequency of the classical laser.

The second term in Eq.~\eqref{sysham} represents the dipole-dipole coupling of the Rb atom Rydberg state $|r_n\rangle$ and the Cs atom Rydberg state $|r^{\prime}\rangle$ to the Rb atom Rydberg state $|p_n\rangle$ and Cs atom Rydberg state $|p^{\prime} \rangle$. Interactions among Rydberg-excited Rb atoms are neglected, because the system is restricted to host no more than a single (photonic or atomic) excitation at a time. The third  term in Eq.~\eqref{sysham} is the small F{\"o}rster energy penalty $\delta$ of the Rydberg product states $|p_{n}\rangle|p^{\prime}\rangle$.

The Hamiltonian of the continuum of field modes impinging on the cavity reads
 \begin{equation}
 H_{B_c} = \hbar \int d\omega \, \omega \, \hat{a}_c^{\dagger}(\omega) \,  \hat{a}_c(\omega),
 \end{equation}
 where the annihilation and creation operators at frequency $\omega$ satisfy the canonical bosonic commutation relation $[\hat{a}_c(\omega),\hat{a}_c^{\dagger}(\omega^{\prime})] = \delta(\omega - \omega^{\prime})$.
The coupling through the cavity mirror between the single cavity mode $\hat{b}$ and the incident field is represented by the interaction Hamiltonian
 \begin{equation}
 H_{\mathrm{int}, B_c} = i \hbar \int d\omega \,  g_c(\omega) \,  \left( \hat{a}_c^{\dagger}(\omega) \, \hat{b}   -  \hat{b}^{\dagger} \, \hat{a}_c(\omega)  \right ),
 \end{equation}
where $g_c(\omega)$ is in general frequency dependent.

The remaining two terms in the Hamiltonian of Eq.~(\ref{HAM}) represent the coupling between the atoms and their environmental bath degrees of freedom. Assuming atomic decay associated with the emission of fluorescence photons, these terms give rise to irretrievable loss of population of the excited states of the system, and they can be described by a loss of norm of the state vector, corresponding to the "no-jump" component in a Monte Carlo wavefunction treatment of the problem \cite{MCWF}.  Without detailing the explicit form of $ H_{B_a}$ and $H_{\mathrm{int}, B_a}$, we expand the "no-jump" state vector as follows (for the case where the Cs atom is initially excited to the Rydberg state $|r'\rangle$),
\begin{align}
\label{2}
|\Psi (t) \rangle &=  \int d\omega \, \phi_c(\omega,t) \, \hat{a}^{\dagger}_{c}(\omega)\, |g^{N},r',\O_b, \O_a\rangle \,  \nonumber \\
&\,\,\,\,\,\,+C_{b}(t)\, \hat{b}^{\dagger}\,|g^{N},r',\O_b, \O_a\rangle \,   \nonumber\\
&\,\,\,\,\,\,+\sum_{m=1}^N C_{em}(t) |g^{N-1},e_{m},r',\O_b, \O_a\rangle \,  \nonumber\\
&\,\,\,\,\,\,+ \sum_{m=1}^N C_{rm}(t)|g^{N-1},r_{m},r',\O_b, \O_a\rangle \,  \nonumber\\
&\,\,\,\,\,\,+ \sum_{m=1}^N C_{pm}(t)|g^{N-1},p_{m},p',\O_b, \O_a\rangle.
\end{align}
Here, $|g^{N}\rangle \equiv |g_1, ..., g_i, ..., g_N \rangle$, while $|g^{N-1},x_{m}\rangle$ denotes the state with all ensemble atoms in the ground state except the $m^{th}$ atom, which is excited to either $|e_{m}\rangle$, $|r_{m}\rangle$ or $|p_{m}\rangle$. $|\O_b\rangle$ denotes the vacuum of the cavity mode and $|\O_a\rangle$ denotes the multimode vacuum state of the field outside the cavity. The Schr{\"o}dinger equation governs the evolution of the state vector amplitudes: $\phi_c(\omega,t)$ on one-photon states with frequency $\omega$ outside the cavity, $C_{em}(t)$, $C_{rm}(t)$ and $C_{pm}(t)$ on atomic excited states and $C_{b}(t)$ on the one-photon state in the cavity. The frequency $\omega$ is defined relative to the cavity frequency $\omega_c$. This implies that $\int d \omega \equiv \int_{-\vartheta}^{+\vartheta} d \omega    =  \int_{\omega_c-\vartheta}^{\omega_c+\vartheta} d \omega^{\prime}$, where $\omega^{\prime} = \omega + \omega_c$ is the real optical frequency of the impinging photon, while $\vartheta \ll \omega_c$ is an appropriate frequency  cutoff.

Assuming spontaneous decay rates $\Gamma_e$, $\Gamma_r$ and $\Gamma_p$ for the Rb excited states (i.e. we neglect collective phenomena like sub- and super-radiance), the state vector amplitudes solve the following system of equations
\begin{align}
&\dot{C}_{em}(t) =  i \Omega_{m}^{\ast}C_{rm}(t) - \, C_{b}(t) \, \mathcal{G}_{m} - \frac{\Gamma_e}{2} C_{em}(t), \label{eq.ce}\\
&\dot{C}_{rm}(t) =  i \Omega_{m}C_{em}(t) - i V_m C_{pm}(t)  - \frac{\Gamma_r}{2} C_{rm}(t), \label{eq.cr}\\
&\dot{C}_{pm}(t) = - i V_m C_{rm}(t) -i \delta C_{pm}(t) - \frac{\Gamma_p}{2} C_{pm}(t), \label{eq.cp}\\
&\, \dot{C}_{b}(t) =   \sum_{n=1}^N   \mathcal{G}^{\ast}_{n} C_{en}(t) - \! \int d\omega \, g_c(\omega) \phi_c(\omega,t), \label{eq.cb} \\
&\,  \dot{\phi}_c(\omega,t) = -i \,\omega \,  \phi_c(\omega,t) + g_c(\omega) \, C_b(t). \label{eq.f1}
\end{align}
The formal solutions of Eq.~(\ref{eq.f1}) is:
\begin{align}
\phi_c(\omega,t) &=  e^{-i\omega t}  \phi_c(\omega,0) + g_c(\omega) \int_{0}^t ds \, e^{-i \omega (t-s)} C_b(s), \label{eq.sf1}
\end{align}
which inserted in Eq.~(\ref{eq.cb}) yields
\begin{equation}
\dot{C}_b(t) = \sum_{n=1}^N  \mathcal{G}^{\ast}_{n} C_{en}(t)   -\frac{\kappa}{2} C_b(t) - \sqrt{\kappa} \, \beta_{\mathrm{in}}(t). \label{sol4}
\end{equation}
Eq.~(\ref{sol4}) with $\kappa=2\pi |g_c(\omega'=\omega_c)|^2$ is obtained in the Markov approximation upon the assumption of a smooth frequency dependence of $g_c(\omega)$. $\beta_{\mathrm{in}}(t) = \frac{1}{\sqrt{2\pi}} \int d\omega \, e^{-i\omega t} \phi_c(\omega,0)$ represents the time dependent arrival of the initial ($t=0$) photon wave packet incident on the cavity.


Eq.~(\ref{eq.f1}) can also be solved so that it matches the shape of the later (reflected) photon wave packet at  $T>t$,
\begin{align}
\phi_c(\omega,t) &=  e^{-i\omega (t-T)}  \phi_c(\omega,T) \nonumber \\
&- \kappa_c(\omega) \int_{t}^T ds \, e^{-i \omega (t-s)} C_b(s), \label{eq.sf3}
\end{align}
yielding instead,
\begin{align}
&\dot{C}_b(t) = \sum_{n=1}^N \left(  \mathcal{G}^{\ast}_{n} C_{en}(t) \right)  + \frac{\kappa}{2} C_b(t) - \sqrt{\kappa} \, \beta_{\mathrm{out}}(t), \label{sol4-out}
\end{align}
where $\beta_{\mathrm{out}}(t) = \frac{1}{\sqrt{2\pi}} \int d\omega \, e^{-i\omega(t-T)}\, \phi_c(\omega,T)$ thus represents the output cavity field.

Subtracting Eqs.~(\ref{sol4}) and Eqs.~(\ref{sol4-out}), we obtain
\begin{equation} \label{inout}
\beta_{\mathrm{out}}(t) = \beta_{\mathrm{in}}(t) + \sqrt{\kappa} \, C_b(t).
\end{equation}
which provides a relation between the input and output field amplitudes. Note that Eq.(\ref{inout}) is equivalent to the corresponding relation between annihilation operators in the input-output theory formalism~\cite{gardiner2004a}.

All the relevant amplitudes solve a coupled set of linear, first order differential equations. By introducing the Fourier transform, we obtain for each frequency $\omega$ a linear set of equations which can be solved analytically. The output field $\beta_{\text{out}}(\omega)$ can be expressed as
\begin{align}
\label{3}
\beta_{\text{out}}(\omega) &=  R(\omega) \,  \beta_{\text{in}}(\omega),
\end{align}
where the complex reflection coefficient,
\begin{align}
R(\omega)= 1{}&{} - \kappa \left\{\frac{\kappa}{2} - i \omega + \sum_{n=1}^N  |\mathcal{G}_n|^2 \left[ \frac{\Gamma_{e}}{2}-i\omega \phantom{+\frac{|\Omega|^2}{\frac{\Gamma_{r}}{2} - i \omega + \frac{|V_{jn}|^2}{\frac{\Gamma_p}{2}+i(\delta-\omega)}}}\right .\right . \nonumber \\
&\left.\left.\qquad\qquad{}+\frac{|\Omega|^2}{\frac{\Gamma_{r}}{2} - i \omega + \frac{|V_{n}|^2}{\frac{\Gamma_p}{2}+i(\delta-\omega)}}\right]^{-1}\right\}^{-1}.\label{6}
\end{align}
readily generalizes to Eq.~\eqref{eq:full_R} in the main text (lifting the single atom assumption, $\{ j \} = 1$).

\end{document}